\documentclass[%
 aip,
 amsmath,amssymb,
 reprint,%
]{revtex4-1}

\usepackage{graphicx}% Include figure files
\usepackage{dcolumn}% Align table columns on decimal point
\usepackage{bm}% bold math
\usepackage[utf8]{inputenc}
\usepackage[T1]{fontenc}
\usepackage{mathptmx}
\usepackage{etoolbox}
\usepackage{mathrsfs}
\usepackage{amsmath}
\makeatletter
\def\@email#1#2{%
 \endgroup
 \patchcmd{\titleblock@produce}
  {\frontmatter@RRAPformat}
  {\frontmatter@RRAPformat{\produce@RRAP{*#1\href{mailto:#2}{#2}}}\frontmatter@RRAPformat}
  {}{}
}%
\makeatother
\begin{document}

\preprint{AIP/123-QED}

\title{Acoustic Phonon-Induced Dephasing in Gallium Nitride Defect-Based Quantum Emitters}
% Force line breaks with \\
\author{Yifei Geng}
  \affiliation{School of Electrical and Computer Engineering, Cornell University, Ithaca, New York 14853, USA.}%Lines break automatically or can be forced with \\
  \email{yg474@cornell.edu}

\date{\today}% It is always \today, today,

\begin{abstract}
GaN defect-based quantum emitters have recently gained attention as promising single-photon sources for quantum information applications. However, dephasing processes---manifested as photoluminescence (PL) linewidth broadening---pose a limitation to photon indistinguishability. In this study, we use a custom-built confocal scanning microscope to examine the temperature-dependent PL spectra of GaN defect quantum emitters integrated with solid immersion lenses, with the goal of elucidating their dephasing mechanisms. Our experimental findings show that at low temperatures, the PL lineshape exhibits a Gaussian profile with a constant, temperature-independent linewidth, consistent with spectral diffusion. As the temperature increases, the PL lineshape evolves into a Lorentzian, and the temperature-dependent linewidth deviates from the common $T^{3}$ law. Considering the Debye temperature of GaN ($\sim$600 K), the experimentally observed temperature-dependent linewidth can be modeled by the quadratic Stark effect modulated by acoustic phonons in defect-rich crystals. Furthermore, this model exhibits a level of accuracy comparable to that of the defect–$E_{2}$(low) optical phonon coupling model previously reported in the literature. Our work reveals the mechanism of acoustic phonon-induced dephasing in GaN defect emitters and demonstrates that both acoustic and optical phonons can contribute to their dephasing. 
\end{abstract}

\maketitle

A single-photon source constitutes a fundamental building block of quantum information technology\cite{aharonovich2016solid}. An ideal single-photon source should exhibit high brightness, high purity, excellent stability, well-defined linear polarization, high photon indistinguishability, room-temperature operation, and compatibility with on-chip photonic integration\cite{geng2024defect,luo2025gan,geng2022decoherence,yi2019topological}. In recent years, defect-based quantum emitters in GaN have attracted significant research interest due to their potential to demonstrate many of the aforementioned desirable properties\cite{berhane2017bright,geng2022temperature,geng20221temperature}. In contrast to NV centers in diamond, GaN defect emitters exhibit strong zero-phonon line emission even at room temperature, and their PL displays pronounced linear polarization\cite{geng2025temperature}. More importantly, GaN is a well-established direct bandgap semiconductor with widespread applications in both electronic and photonic devices\cite{wasisto2019beyond,zhu2022characteristics,ma2019review,chen2017gan}, offering a significant advantage for the on-chip integration of its defect-based quantum emitters. Many physical properties of GaN defect quantum emitters have been gradually characterized. These emitters have been shown to exhibit emission spectra in the 600-700 nm range\cite{geng2023dephasing}, excited-state lifetimes on the order of nanoseconds\cite{berhane2018photophysics}, and optical dipoles oriented nearly perpendicular to the c-axis of the wurtzite crystal\cite{geng2023optical}. Additionally, their spin properties\cite{luo2024room,luo20241room,luo2024data,luo2023room} and ultrafast spectral diffusion characteristics\cite{geng2023ultrafast} have also been reported. However, as defect-based single-photon emitters in the early stages of development, their nature remains not fully understood. Considering the symmetry of their optical dipoles and recent findings from first-principles calculations, impurity defects and impurity–vacancy pair complexes have emerged as promising candidates\cite{yuan2023gan,geng2023optical}.

Among all their properties of interest, photon indistinguishability is particularly significant, as it enables quantum interference, which is essential for entanglement generation, quantum logic operations, and high-fidelity quantum communication\cite{santori2002indistinguishable}. An ideal single-photon emitter can generally be modeled as a two-level system. In the absence of environmental interactions, the emission spectrum can be approximated as a Dirac delta function, with an extremely narrow linewidth primarily determined by the excited-state lifetime and the energy-time uncertainty principle. However, in realistic environments, single-photon emitters inevitably experience various dephasing processes, which lead to broadening of the emission linewidth. In solid-state materials, defect-phonon interactions and spectral diffusion are the primary dephasing mechanisms. Spectral diffusion is known to contribute to a temperature-independent Gaussian lineshape. Defect–phonon interactions contribute to the temperature-dependent broadening of the emission linewidth, and many different physical models have been proposed to describe this behavior. For instance, the PL linewidth $T^{3}$ dependence observed in AlN, SiC, and hBN defect single-photon emitters is attributed to dephasing induced by acoustic phonons in crystals with a high defect density\cite{xue2020single,sontheimer2017photodynamics,lienhard2016bright,hizhnyakov1999optical}. The dynamic Jahn-Teller effect in the excited state of NV centers in diamond gives rise to a $T^{5}$ dependence\cite{fu2009observation,abtew2011dynamic}. Quadratic coupling with acoustic phonons\cite{silsbee1962thermal,hizhnyakov2002zero} results in a temperature-dependent linewidth that follows $T^{7}$.

\begin{figure*}[htb]
\includegraphics[width=0.88\textwidth]{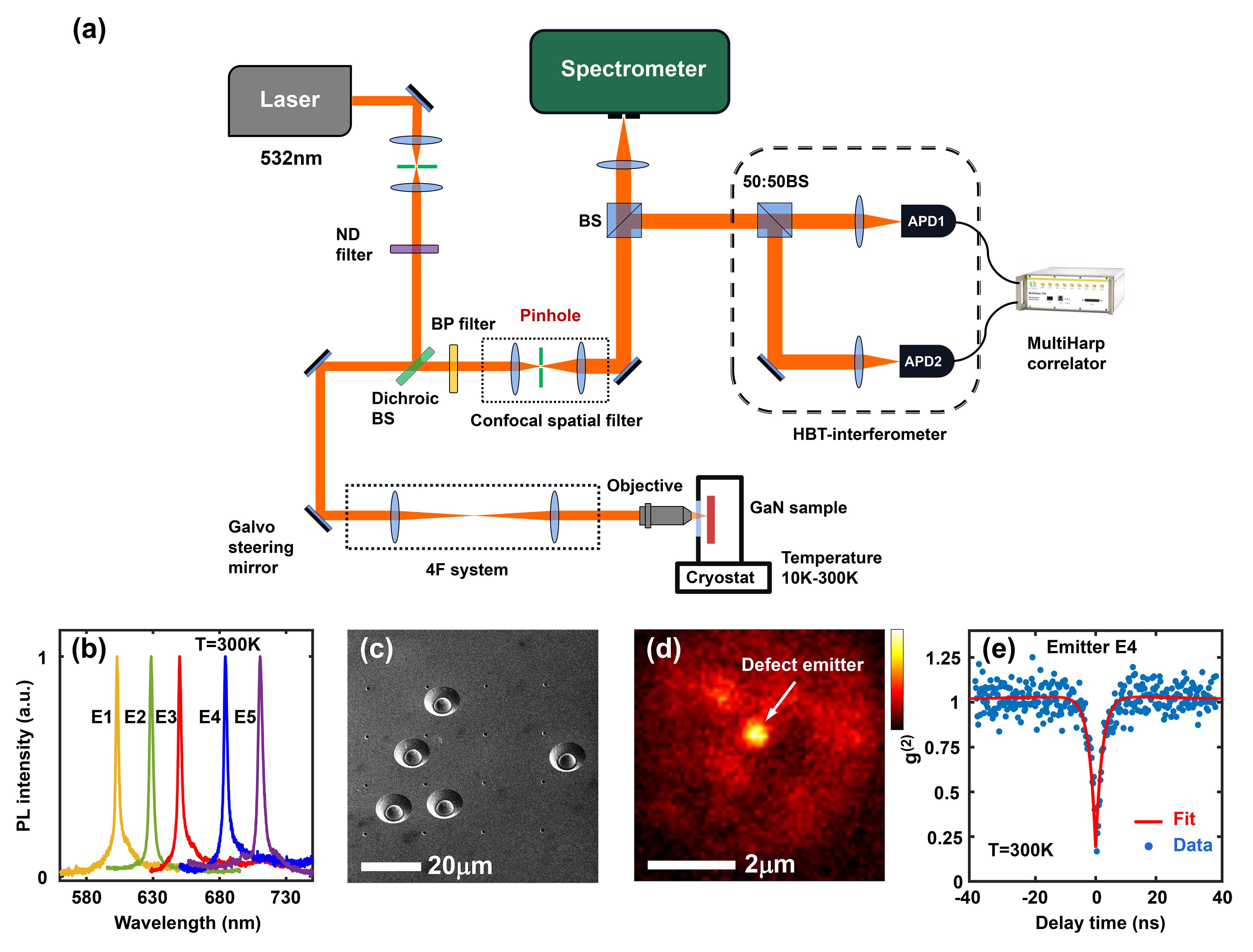}% Here is how to import EPS art
\caption{\label{fig1} (a) Schematic layout of the custom-built confocal scanning microscope, equipped with a spectrometer and an HBT interferometer. (b) Room-temperature PL spectra of five defect emitters, labeled E1 to E5. (c) Representative SEM image of five fabricated solid immersion lenses (SILs). Each SIL is a hemisphere with a diameter of 5 $\mu m$. (d) Spatial PL map of a defect emitter inside a SIL. The color bar represents the normalized PL intensity. (e) Measured $g^{(2)}(\tau)$ function of emitter E4 at room temperature. Reprinted with permission from Ref [14]. Copyright 2023, The Authors.\cite{geng2023dephasing}}
\end{figure*}

However, the temperature-dependent PL linewidth of GaN defect emitters does not follow the commonly observed power laws of $T^{3}$, $T^{5}$, or $T^{7}$. In 2023, Geng et al. proposed a model based on defect-optical phonon coupling, which successfully explains the experimental results\cite{geng2023dephasing}. Defect-optical phonon coupling is generally not considered a significant dephasing process at temperatures well below room temperature, due to the high energy of optical phonons. What makes GaN unique is its low-energy (18 meV) Raman-active optical phonon mode, $E_{2}(low)$, at the center of the Brillouin zone, which likely contributes to the dephasing process. This defect-optical phonon coupling model has successfully explained the deviation of GaN defect emitters from the $T^{3}$ law. However, it raises a further question: Given the widespread presence of low-energy acoustic phonons in crystals, it is puzzling why they do not contribute to dephasing in GaN defect emitters.

In this work, we investigate the temperature-dependent PL spectrum of GaN defect emitters integrated with solid immersion lenses and examine the role of acoustic phonons in their dephasing process. Our experimental results show that, at low temperatures, the PL spectral lineshape is Gaussian, with the linewidth saturating at $\sim$1 meV. As the temperature increases, the PL spectrum gradually evolves into a Lorentzian lineshape. Given that the Debye temperature of GaN is only 600 K, the experimentally observed temperature-dependent linewidth of GaN single-photon emitters can be explained by the quadratic Stark effect modulated by acoustic phonons in defect-rich crystals. Furthermore, this model proves to be nearly as effective as the defect-$E_{2}(low)$ optical phonon coupling model proposed in earlier work, providing a compelling explanation for the deviation of the PL linewidth of GaN defect emitters from the $T^{3}$ dependence. Our work reveals the mechanism of acoustic phonon-induced dephasing in GaN defect emitters, demonstrating that both acoustic and optical phonons can contribute to the dephasing process in GaN.

We begin by presenting the experimental setup and the results. As shown in Fig.\ref{fig1} (a), a custom-built confocal scanning microscope equipped with a Hanbury Brown and Twiss (HBT) interferometer was used for the investigation. A 532 nm laser was used to excite the GaN defect emitter, with a galvo steering mirror and a 4F system employed to scan the laser. The GaN sample was mounted inside a helium-flow cryostat with temperature control ranging from 10 K to 300 K. The PL light from the sample was collected by an objective lens (NA = 0.7) with a correction collar through the cryostat window. A confocal spatial filter was placed in the collection path to ensure a high SNR. A spectrometer and an HBT interferometer were used to measure the PL spectrum and the second-order correlation function $g^{(2)}(\tau)$.

Our sample is a 4 $\mu m$ thick HVPE-grown semi-insulating GaN layer on a sapphire substrate. Fig.\ref{fig1} (b) presents the PL spectra of five defect emitters, labeled E1 to E5, at room temperature. GaN is a well-known high-refractive-index material, with a refractive index of up to 2.4 in the visible wavelength range. This is highly advantageous for integrated photonics due to its ability to confine light modes more effectively; however, it presents challenges in efficiently collecting light emitted from defect emitters embedded within. The PL light emitted from defect emitters undergoes total internal reflection at the GaN–air interface, significantly limiting the amount of light that can be collected. To address this issue, we fabricated a solid immersion lens (SIL) above each defect emitter using focused ion beam (FIB) etching. The SIL is a hemisphere with a diameter of 5 $\mu m$, as shown in the SEM image in Fig.\ref{fig1} (c). Fig.\ref{fig1} (d) shows a spatial PL map of a defect emitter inside a SIL. When the defect emitter is close to the center of the SIL, the collection efficiency can be enhanced by a factor of up to $\sim$5. Fig.\ref{fig1} (e) presents the measured $g^{(2)}(\tau)$ function of the defect emitter E4 (shown in Fig.\ref{fig1} (b), with a PL center wavelength of 684.5 nm), and $g^{(2)}(\tau=0)$ is below 0.5, confirming its identity as a single-photon emitter. Further details can be found in the previous work\cite{geng2023dephasing,geng2024defect}.

\begin{figure}
\includegraphics[width=1\columnwidth]{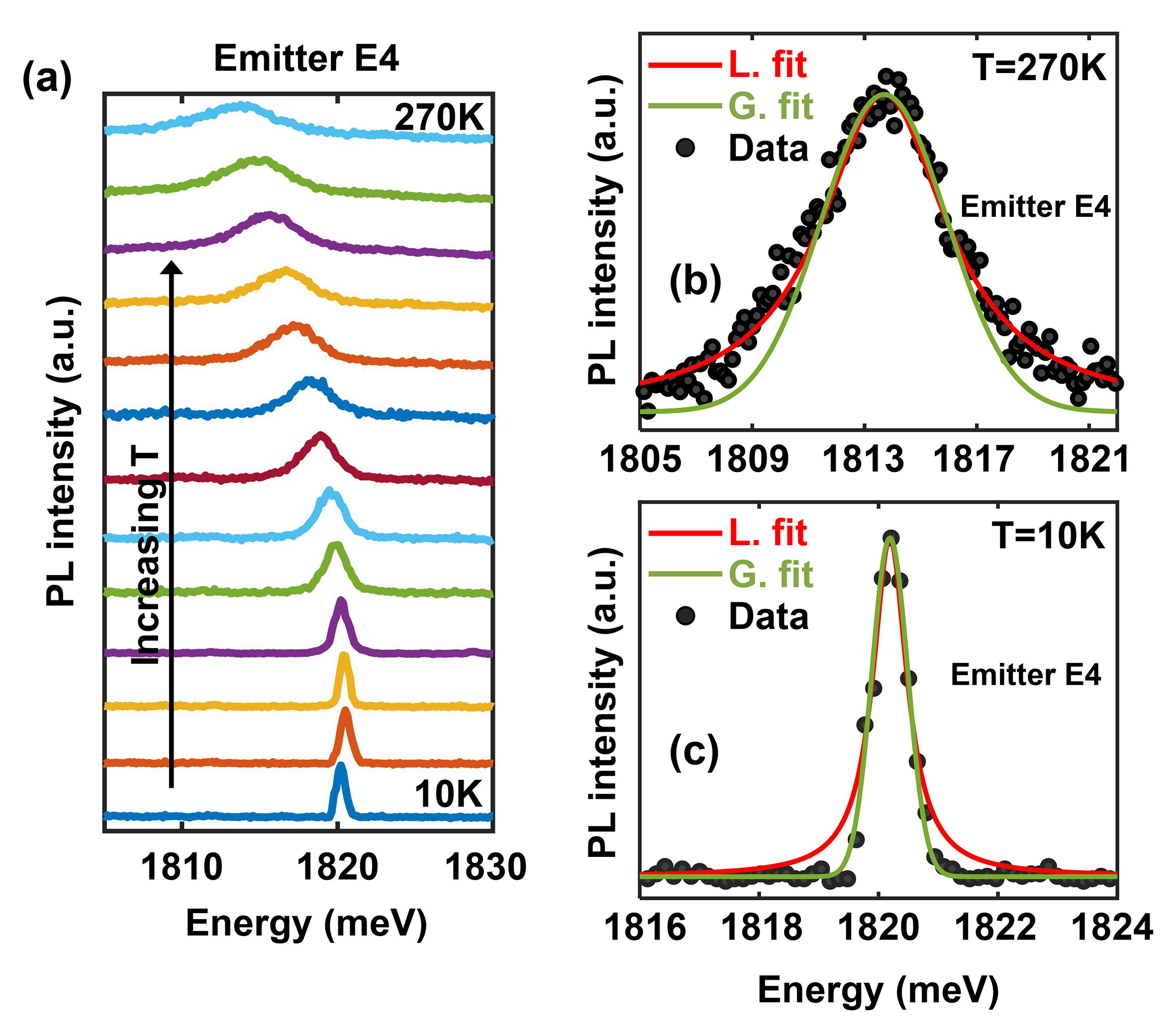}% Here is how to import EPS art
\caption{\label{fig2} 
(a) Temperature-dependent PL spectra of emitter E4 from 10 K to 270 K, in 20 K increments. (b) Measured PL data (black dots) at 270 K can be better fitted using a Lorentzian function (red curve). (c) Measured PL data (black dots) at 10 K can be better fitted using a Gaussian function (green curve). Reprinted with permission from Ref [14]. Copyright 2023, The Authors.\cite{geng2023dephasing}}
\end{figure}

In the following, we use the defect emitter E4 as an example to investigate its temperature-dependent PL spectrum. Fig.\ref{fig2} (a) shows its temperature-dependent PL spectra, recorded from 10 K to 270 K in 20 K increments. Firstly, we observe that as the temperature increases, the PL spectrum exhibits a redshift, with the PL center energy shifting from 1820.2 meV at 10 K to 1813.5 meV at 270 K, accompanied by a gradual broadening of the linewidth. Secondly, we observe that the PL spectral lineshape evolves with temperature. At low temperature (10 K), the lineshape is Gaussian, as shown in Fig.\ref{fig2} (c), with a linewidth of 0.72 meV. At high temperature (270 K), the lineshape transitions to Lorentzian, with a linewidth of 6.82 meV, as shown in Fig.\ref{fig2} (b). This suggests that at least two distinct mechanisms contribute to the dephasing process. In fact, the Gaussian lineshape at low temperatures is well known to result from spectral diffusion, which is caused by local electric field fluctuations, such as those induced by charge carrier creation and annihilation due to the excitation laser. Spectral diffusion-induced Gaussian lineshapes have been observed in many single-photon emitters of different materials, with the Gaussian linewidth being temperature-independent, commonly referred to as "inhomogeneous broadening". In contrast, the Lorentzian lineshape, which broadens with increasing temperature, suggests the involvement of phonons in the dephasing process. Therefore, the accurate lineshape should be a Voigt function, which is the convolution of a Gaussian and a Lorentzian,
\begin{equation}
    V(\omega;\sigma,\gamma) \propto \int_{-\infty}^{+\infty}G(\omega^{'};\sigma)L(\omega-\omega^{'};\gamma)d\omega^{'}
    \label{eq1}
\end{equation}
Here, $G(\omega;\sigma)$ and $L(\omega;\gamma)$ are Gaussian and Lorentzian functions with FWHM linewidth given by $f_{G} = 2\sigma\sqrt{2\ln2}$ and $f_{L} = 2\gamma$, respectively. The FWHM $f_{V}$ of the Voigt function can be written as (not a simple algebraic addition),
\begin{equation}
    f_V=0.5346f_L+\sqrt{0.2166 f_{L}^{2}+f_{G}^{2}} \label{eq2}
\end{equation}

By fitting the measured PL spectra with a Voigt function, the temperature-dependent total linewidth (FWHM) and its Gaussian and Lorentzian components can be extracted. The Gaussian component $f_{G}$ remains constant at 0.72 meV, while the Lorentzian component $f_{L}$ exhibits a clear temperature dependence. As shown in Fig.\ref{fig4}, it does not follow a power law, such as $T^{3}$. Previous work has suggested defect–optical phonon coupling, specifically the $E_{2}(low)$ mode at the Brillouin zone center, as a possible explanation for this deviation. However, due to the unique properties of GaN, this is not the only model capable of explaining the experimental results. In the following discussion, we show that in defect-rich crystals---given GaN’s relatively low Debye temperature of 600 K---the quadratic Stark effect modulated by acoustic phonons can also successfully account for the experimental results and deviation from the $T^{3}$ law.

\begin{figure*}
\includegraphics[width=0.7\textwidth]{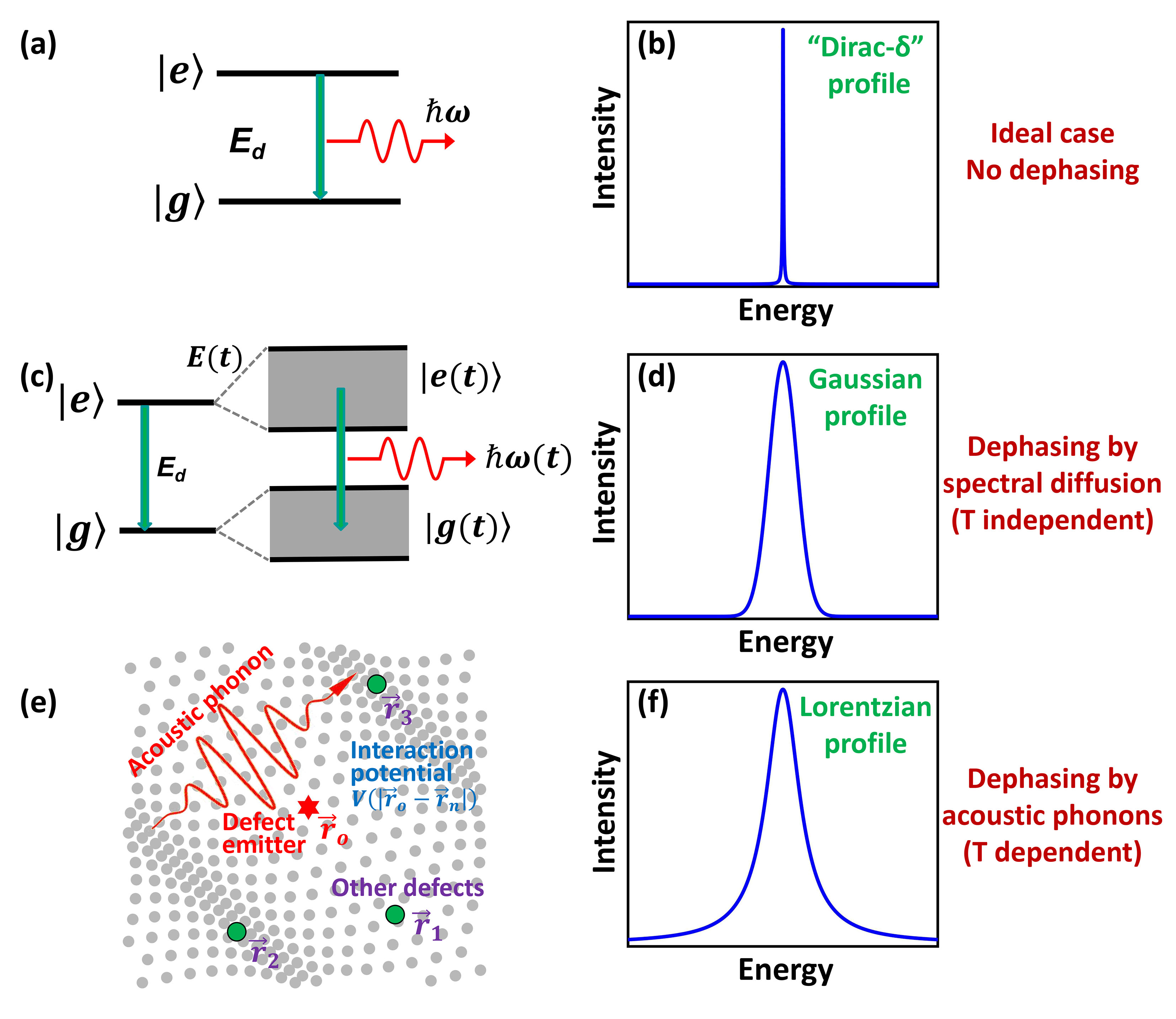}% Here is how to import EPS art
\caption{\label{fig3} 
(a) Ideal two-level system with ground state $|g\rangle$ and excited state $|e\rangle$. (b) In the absence of dephasing, the ideal spectral density function is a Dirac delta profile. (c) When the two-level system is exposed to a (slowly) time-varying external electric field $E(t)$, its energy---and consequently the energy of the emitted photon---is modulated by the field. (d) The emitted photon energy randomly fluctuates under a Gaussian envelope, resulting in a temperature-independent Gaussian spectral density profile. This is known as spectral diffusion. (e) In a defect-rich crystal, if the defect emitter is surrounded by other charged defects, their distances, and thus their interactions, are modulated by acoustic phonons. This modulation leads to a quadratic Stark effect that induces dephasing. (f) The acoustic-phonon-modulated quadratic Stark effect leads to a Lorentzian spectral profile, with the linewidth becoming temperature-dependent as the acoustic phonon occupation number $n(\omega)$ appears in Eq.\ref{eq8}.}
\end{figure*}

As shown in Fig.\ref{fig3}, for an ideal two-level system with ground state $|g\rangle$ and excited state $|e\rangle$, the spectral density function would be a Dirac delta function in the absence of any dephasing. However, when the system is subject to an external time-varying electric field and coupled to surrounding charged defects with acoustic phonon modulation, its spectral density function experiences Gaussian broadening due to spectral diffusion and Lorentzian broadening due to phonon-induced dephasing. We consider an ideal two-level system subject to these two distinct dephasing mechanisms. The corresponding Hamiltonian is given by:
\begin{eqnarray}\label{eq3}
  & &H = E_{d}d^{\dagger}d + \sum_{\vec{k}} \hbar \omega_{\vec{k}} a_{\vec{k}}^{\dagger}a_{\vec{k}} \\
  & & + \alpha E(t)  d^{\dagger}d \nonumber + d^\dagger d\sum_{n,m}B_{jk}\Delta E_n^j\Delta E_m^k
\end{eqnarray}
Here, $d$ and $a_{\vec{k}}$ are the annihilation operators for the defect electron and acoustic phonon states, respectively. The first line represents the unperturbed Hamiltonian $H_{0}$, while the second line denotes the interaction Hamiltonian $H^{\prime}$. The first term in $H^{\prime}$ represents the coupling between the defect and an external time-varying electric field $E(t)$, as illustrated in Fig.\ref{fig3} (c). We assume that $\langle E(t) \rangle = 0$ and $\langle E(t) E(t')\rangle = E_{0}^{2} e^{-\lambda |t-t'|}$, where $\lambda^{-1}$ is the field correlation time, which we assume to be much longer than any other time scale in the problem. Indeed, previous experimental studies have shown that the correlation time of the external field is on the microsecond scale, significantly exceeding the nanosecond-scale characteristic lifetime of defect emitters in GaN\cite{geng2023ultrafast,berhane2018photophysics}. When the two-level system is exposed to this time-varying electric field, its energy---and consequently the energy of the emitted photon---is modulated by the field. This modulation leads to random spectral jumps within a Gaussian envelope in the emission spectrum, a characteristic signature of spectral diffusion, as shown in Fig.\ref{fig3} (d).

The second term in $H^{\prime}$ represents the coupling between the defect emitter and nearby charged defects in a defect-rich crystal, mediated by acoustic phonon modulation (quadratic Stark effect). As shown in Fig.\ref{fig3} (e), in a defect-rich crystal, when the defect emitter (represented by a red star, with position vector $\vec{r}_0$) is surrounded by other charged defects (green dots, with position vectors $\vec{r}_n$), the interaction potential is $V\left(\left|\vec{r}_0-\vec{r}_n\right|\right)$, and the interaction field is given by $E_n^j=-\partial_jV\left(|\vec{r}_0-\vec{r}_n|\right)$. Acoustic phonons in the crystal modify the relative distance between the charged defects and the emitter, resulting in a change in the interaction field,
\begin{equation}
\Delta E_n^j  =-\partial_\ell\partial_jV\left(\left|\vec{r}_0-\vec{r}_n\right|\right)\left(u^\ell\left(\vec{r}_0\right)-u^\ell\left(\vec{r}_n\right)\right) 
\label{eq4}
\end{equation}
Here, $\vec{u}$ is the acoustic photon amplitude, given by:
\begin{equation}
\vec{u}\left(\vec{r}_j\right)=\sum_{\vec{k}}\sqrt{\frac{\hbar}{2NM_j\omega_{\vec{k}}}}\hat{e}_{\vec{k}}
\begin{pmatrix}
a_{\vec{k}}+a_{-\vec{k}}^\dagger
\end{pmatrix} e^{i\vec{k} \cdot \vec{r}_j}
\label{eq5}
\end{equation}

By solving the Schrodinger equation simultaneously with Eqs. \ref{eq3}, \ref{eq4}, and \ref{eq5}, the time evolution of the state vector and the spectral density function can be obtained (see the Supplementary Material for the derivation),
\begin{equation}\label{eq6}
\left|\psi\left(t\right)\right\rangle=e^{-iH_{0} t/\hbar}\pmb{\mathcal{T}}\left\{e^{-i\int\limits_{0}^{t} H_{I}^{\prime}(t^{\prime})dt^{\prime}/\hbar}\right\}|\psi\left(t=0\right)\rangle
\end{equation}

\begin{equation}\label{eq7}
  S(\omega) \approx \int dt \, e^{-i(\omega - E_{d}/\hbar) t} \, e^{-\sigma^{2}t^{2}/2} e^{-\gamma \, |t|} 
\end{equation}
It is evident from Eq.\ref{eq7} that the spectral density function is a convolution of a Gaussian profile and a Lorentzian profile in $\omega$-domain as seen previously in Eq.\ref{eq1}. For the Gaussian function, $\sigma = |\alpha E_{0}/\hbar|$ (a temperature-independent constant) and for the Lorentzian function,
\begin{equation}
\gamma\propto\overset{\omega_D}{\underset{0}{\operatorname*{\operatorname*{\int}}}}d\omega\; \omega^2\mathrm{~}n(\omega)(n(\omega)+1)
\label{eq8}
\end{equation}
Here, $n(\omega)$ is the Bose-Einstein occupation function, and the Debye frequency is given by $\omega_{D}=k_{B}\theta_{D} / \hbar$, where $k_{B}$ is the Boltzmann constant and $\theta_{D}$ is the Debye temperature.

\begin{figure}
\includegraphics[width=0.85\columnwidth]{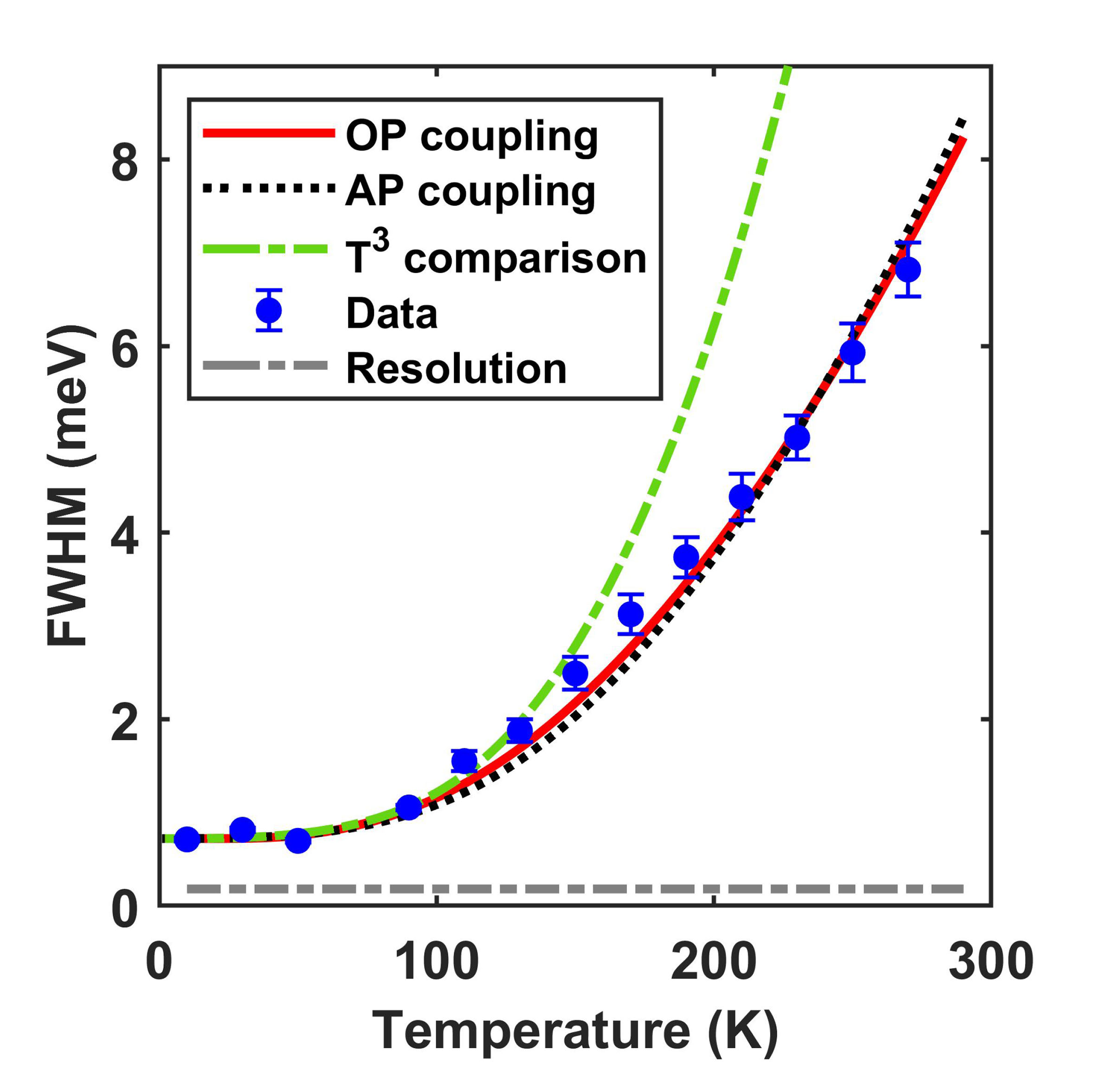}% Here is how to import EPS art
\caption{\label{fig4} 
The temperature-dependent linewidth data (blue dots) and fits using different models. The red curve represents the optical phonon coupling model proposed in earlier work\cite{geng2023dephasing}, the black dotted line corresponds to the acoustic phonon-induced dephasing model in this work using a finite Debye temperature of 600 K, and the green dashed line is the $T^{3}$ comparison.}
\end{figure}

The integral in the form of Eq.\ref{eq8} is quite interesting. If the Debye temperature is much higher than the temperature range of interest, the upper limit of the integration can be approximated as positive infinity, yielding an analytical expression for the integral $\overset{+\infty}{\underset{0}{\operatorname*{\operatorname*{\int}}}} d\omega\;\omega^2\mathrm{~}n(\omega)(n(\omega)+1)=\frac{\pi^2}{6}\left(k_BT/\hbar\right)^3$. It shows that the Lorentzian linewidth exhibits a $T^{3}$ dependence on temperature. This $T^{3}$ dependence may universally apply at low temperatures, particularly in defect-rich crystals, regardless of the specific material. It indicates a more efficient dephasing mechanism compared to the dynamic Jahn-Teller effect (which gives a $T^{5}$ dependence) and defect-acoustic phonon quadratic coupling (which gives a $T^{7}$ dependence). In fact, defect emitters in defect-rich crystals, such as AlN ($\theta_{D}$ $\sim$1150 K \cite{AlNDebye}), SiC ($\theta_{D}$ $\sim$1300 K \cite{SiCDebye}), and diamond ($\theta_{D}$ $\sim$1900 K \cite{schoening1969x} or $\sim$2200 K \cite{mukherjee1967debye}), which have sufficiently high Debye temperatures, are more likely to exhibit this $T^{3}$ behavior. GaN, however, with a Debye temperature of only $\sim$600 K\cite{GaNDebye,chen1999structure}, may not allow the upper limit of the integral in Eq.\ref{eq8} to be approximated as infinity. Consequently, the temperature-dependent linewidth does not follow the $T^{3}$ law as the temperature increases. For a finite Debye temperature as the upper limit of integration, Eq.\ref{eq8} does not have an analytical solution and must be evaluated numerically.

As demonstrated in Fig.\ref{fig4}, the temperature-dependent total linewidth data (blue dots) are shown along with different fitting curves. The dashed green curve represents the $T^{3}$ model, which can fit the experimental data at very low temperatures (below 120 K), where the temperature is much lower than the Debye temperature, allowing the integral upper limit to be approximated as infinity. However, this fit rapidly deviates as the temperature increases. The black dotted curve represents the acoustic phonon-induced dephasing model presented in this work, where Eq.\ref{eq8} is solved numerically using a Debye temperature of 600 K. It fits the experimental data nearly as well as the red curve, which represents the defect-$E_{2}(low)$ optical phonon coupling model proposed in previous work\cite{geng2023dephasing}.

GaN is a unique material that not only possesses a Raman-active low-energy (18 meV) optical phonon mode ($E_{2}(low)$) at the Brillouin zone center---enabling optical phonons to participate in the dephasing process---but also has a relatively low Debye temperature, which makes the commonly observed acoustic phonon-induced $T^{3}$ law inapplicable. Although both acoustic phonon and optical phonon induced dephasing mechanisms can explain the experimental data with comparable accuracy, their physical origins are fundamentally different. Optical phonon coupling is regarded as an intrinsic property of the defect emitter, whereas acoustic phonon coupling depends on the local environment---specifically, the presence of nearby charged defects in the GaN crystal. Clearly, further experimental and theoretical investigations---particularly on emitters in high-purity, low-defect-density GaN---are necessary to fully elucidate the underlying dephasing mechanisms.

In summary, we have employed a custom-built confocal scanning microscope to study the temperature-dependent PL spectrum of GaN defect quantum emitters integrated with SILs. Our experimental results show that at low temperatures, the PL lineshape is Gaussian, and the linewidth saturates at a temperature-independent constant, consistent with spectral diffusion. As the temperature increases, the PL lineshape evolves into a Lorentzian profile. We have explained the deviation of the temperature-dependent linewidth from the commonly observed $T^{3}$ behavior. We find that both optical and acoustic phonons can contribute to the dephasing process, resulting in nearly identical lineshape and linewidth broadening. We hope that this work could inspire further experimental and theoretical studies on GaN defect emitters.

\begin{acknowledgments}
The author thanks Professor Farhan Rana for helpful discussions. This work was supported by the Cornell Center for Materials Research through the NSF MRSEC program (DMR-1719875), the NSF RAISE:TAQS program (ECCS-1839196), and the NSF EAGER program (CMMI-2240267).
\end{acknowledgments}

\nocite{*}
\bibliography{aipsamp}% Produces the bibliography via BibTeX.

\end{document}